% ------------------------------------------------------------------- 
%	Singular perturbation expansions for 
%		H=d^2/dx^2+Bx^2+lambda/x^alpha
%	and
%		H=d^2/dx^2+Bx^2+A/x^2+lambda/x^alpha
% -------------------------------------------------------------------
% Joint paper Saad-Hall-Keviczky
% 
% spe.tex   22 November 2002 [7 August 2003][30 Sept 2003] 
%
% -------------------------------------------------------------------
\def\ptitle{\tiny Perturbation expansions for a class of singular potentials}
% -------------------------------------------------------------------
%  generic unix 12 fonts (lower case names) with no magstep
% --------------------------------------------------------------------
\font\tr=cmr12                          % Our default
\font\bf=cmbx12                         % Redefinition
                         % Redefinition
\font\it=cmti12                         % Redefinition
\font\trbig=cmbx12 scaled 1500          % Main Title
\font\th=cmr12                          % Theorems                       
\font\tiny=cmr10                        % Running title
% --------------------------------------------------------------------
\output={\shipout\vbox{\makeheadline
                                      \ifnum\the\pageno>1 {\hrule}  \fi 
                                      {\pagebody}   
                                      \makefootline}
                   \advancepageno}

\headline{\noindent {\ifnum\the\pageno>1 
                                   {\tiny \ptitle\hfil
page~\the\pageno}\fi}}
\footline{}
% ---------------------------------------------------------------------

\tr 
%--------------------------------------------------------------------
    % bra ket:  math mode (to replace angle)
    %   ket  >
\def\nll{\hfil\break\noindent}  % new line after displayed equations
\def\nl{\noindent}             % noindent

\def\ppl#1{{\noindent\leftskip 9 cm #1\vskip 0 pt}} % Preprint line 

 % bra < math mode
 % ket > math mode
\def\hi#1#2{$#1$\kern -2pt-#2} % hyphen \hi{N}{body} = N-body
\def\hy#1#2{#1-\kern -2pt$#2$} % hyphen hy{large}{N} = large-N
%--------------------------------------------------------------------
\def\dbox#1{\hbox{\vrule % Open box size 2#1 (Abrahams p 273) 
\vbox{\hrule \vskip #1\hbox{\hskip #1\vbox{\hsize=#1}\hskip #1}\vskip #1 
\hrule}\vrule}} 
\def\qed{\hfill \dbox{0.05true in}} % QED 
 % SQUARE 

%--------------------------------------------------------------------
% SPACING
% -------------------------------------------------------------------
\baselineskip 15 true pt  % draft 15 
\parskip=0pt plus 5pt 
\parindent 0.25in
\hsize 6.0 true in 
\hoffset 0.25 true in 
% 6 in width with 1.25 in margins default = (6.5, 0)
\emergencystretch=0.6 in                 % TEXBook p 107 : allows h-space 
\vfuzz 0.4 in                            % page-length flexibility
\hfuzz  0.4 in                           % line-length flexibility
\vglue 0.1true in
\mathsurround=2pt                        % Default is 2pt
\topskip=24pt                            % Default is 10pt
% ---------------------------------------------------------------------
%  References
% ---------------------------------------------------------------------
\newcount\zz  \zz=0  % switch for printing references
\newcount\q   %  reference number
\newcount\qq    \qq=0  % starting reference number-1   (usually zero)

\def\pref#1#2#3#4#5{\frenchspacing \global \advance \q by 1     % paper reference
    \edef#1{\the\q}{\ifnum \zz=1{\item{$^{\the\q}$}{#2}{\bf #3},{ #4.}{~#5}\medskip} \fi}}

\def\bref #1#2#3#4#5{\frenchspacing \global \advance \q by 1     % book reference
    \edef#1{\the\q}
    {\ifnum \zz=1 { %
       \item{$^{\the\q}$} 
       {#2}, {\it #3} {(#4).}{~#5}\medskip} \fi}}

\def\gref #1#2{\frenchspacing \global \advance \q by 1  % general reference
    \edef#1{\the\q}
    {\ifnum \zz=1 { %
       \item{$^{\the\q}$} 
       {#2.}\medskip} \fi}}

 \def\sref #1{#1}

\def\references#1{\zz=#1
   \parskip=2pt plus 1pt   % default is 0pt plus 1pt       
   {\ifnum \zz=1 {\noindent \bf References \medskip} \fi} \q=\qq
%--------------------------------------------------------------------
\pref{\harr}{E. M. Harrell, Ann. Phys. }{105}{ 379 (1977)}{}
%-----------------------------------------------------------------------------
\pref{\green}{W. M. Greenlee, Bull. Amer. Math. Soc. }{82}{341 (1976)}{}
%-------------------------------------------------------------------------
\pref{\hala}{R. Hall, N. Saad and A. von Keviczky, J. Math. Phys. }{39}{6345 (1998)}{}
%-----------------------------------------------------------------------------
\pref{\halb}{R. Hall and N. Saad, J. Phys. A: Math. Gen. }{33}{569 (2000)}{}
%-----------------------------------------------------------------------------
\pref{\halc}{R. Hall and N. Saad, J. Phys. A: Math. Gen. }{33}{5531 (2000)}{}
%-----------------------------------------------------------------------------
\pref{\hald}{R. Hall, N. Saad and A. von Keviczky, J. Phys. A: Math. Gen. }{34}{1169 (2001)}{}
%-----------------------------------------------------------------------------
\pref{\hale}{R. Hall, N. Saad and A. von Keviczky, J. Math. Phys. }{43}{94 (2002)}{}
%------------------------------------------------------------------------------------
\pref{\muod}{Omar Mustafa and Maen Odeh, J. Phys. B: At. Mol. Opt. Phys. }{32} {3055 (1999)  }{}
%------------------------------------------------------------------------------------
\pref{\muod}{Omar Mustafa and Maen Odeh, J. Phys. A: Math. Gen. }{33} {5207 (2000)}{}
%------------------------------------------------------------------------------------
\gref{\jski}{J. Skibi\'nski, e-print quant-ph/0007059}
%---------------------------------------------------------------------------------------
\pref{\klaa}{J. R. Klauder, Acta Phys. Austriaca Suppl. }{XI}{341 (1973)}{}
%-----------------------------------------------------------------------------
\pref{\Klab}{J. R. Klauder, Phys. lett. B }{47}{523 (1973)}{}
%-----------------------------------------------------------------------------
\pref{\klac}{J. R. Klauder, Science }{199}{735 (1978)}{}
%-----------------------------------------------------------------------------
\pref{\eks}{H. Ezawa, J. R. Klauder, and L. A. Shepp, J. Math. Phys. }{16}{783 (1975)}{}
%-----------------------------------------------------------------------------
\pref{\simon}{B. Simon, J. Functional Analysis }{14}{295 (1973)}{} 
%-----------------------------------------------------------------------------
\pref{\deh}{B. DeFacio and C. L. Hammer, J. Math. Phys. }{15}{1071 (1974)}{}
%-----------------------------------------------------------------------------
\pref{\kata}{T. Kato, Prog. Theor. Phys. }{4}{514 (1949)}{}
\pref{\katb}{T. Kato, Prog. Theor. Phys. }{5}{95 (1950)}{}
\pref{\katc}{T. Kato, Prog. Theor. Phys. }{5}{207 (1950)}{}
%-----------------------------------------------------------------------------
\pref{\detw}{L. C. Detwiler and J. R. Klauder, Phys. Rev. D }{11}{1436 (1975)}{}
%----------------------------------------------------------------------------
\bref{\luk}{Yudell L. Luke}{The Special Functions and their Approximation, Vol. I}{Academic Press, 1969}{pp. 43}
%-----------------------------------------------------------------------------
\pref{\kato}
{T. Kato, J. Phys. Soc. Japan }{4}{334 (1949)}{}
%-----------------------------------------------------------------------------
\pref{\harrell2}
{Evans M. Harrell II, Proc. Amerc. Math. Society }{69}{271 (1978)}{Theorem 2}
%-----------------------------------------------------------------------------
\pref{\gtem}
{G. Temple, Proc. Lond. Math. Soc. }{29}{257 (1928)}{}
%-----------------------------------------------------------------------------
\pref{\tem2}
{G. Temple, Proc. Roy. Soc. }{119}{ 276 (1928)}{}
%-----------------------------------------------------------------------------
\pref{\hayes}
{D. T. Hayes, Canadian J. Phys. }{49}{218 (1971)}{}
%-----------------------------------------------------------------------------
\pref{\zno}
{M. Znojil, Physics Letters A }{164}{ 138 (1992)}{}
\pref{\znoj}
{M. Znojil, J. Math. Phys. }{30}{ 23 (1989)}{}
%-----------------------------------------------------------------------------
\pref{\anlk}
{V. C. Aguilera-Navarro and E. Ley Koo, Int. J. Theor. Phys. }{36}{157 (1997)}{}
%-----------------------------------------------------------------------------
\pref{\pease}
{P. Chang and Chen-Shiung Hsue, Phys. Rev. A }{49}{4448 (1994)}{}
%-----------------------------------------------------------------------------
\pref{\acn}
{V. C. Aguilera-Navarro, A. L. Coelho and N. Uttah, Phys. Rev. A }{49}{1477 (1994)}{}
%-----------------------------------------------------------------------------
\pref{\halh}
{R. Hall and N. Saad, Canad. J. Phys. }{73}{493 (1995)}{}
%-----------------------------------------------------------------------------
\pref{\acn}
{V. C. Aguilera-Navarro, Francisco M. Fern\'andez, R. Guardiola and J. Ros, J. Phys. A: Math. Gen. }{25}{6379 (1992)}{}
%----------------------------------------------------------------------------
\pref{\jam}
{M. J. Jamieson, J. Phys. B: Mol. Phys. }{16} {L391 (1983)}{}

%-----------------------------------------------------------------------------
\bref{\as1}
{M. Abramowitz and I. A. Stegan} {Handbook of mathematical functions}{Dover Publications, Inc., New York, 9$^{th}$ Edition (1970)}{pp. 507, formula 13.4.12}
%-----------------------------------------------------------------------------
}% end of ref list

\references{0}    % Initialization of reference numbers
% ------------------------------------------------------------------ end our ref.tex

% ----------------------------
% Preprint list
% ----------------------------
\ppl{CUQM-100}

\ppl{math-ph/0309067}

\ppl{September 2003}
%-------------------------------------------------------------------
%Title Page 
%-------------------------------------------------------------------
%-------------------------------------------------------------------
%Title Page 
%-------------------------------------------------------------------
\vskip 0.5 true in
\centerline{\bf\trbig Perturbation expansions for a class of}
\vskip 0.2 true in
\centerline{\bf\trbig singular potentials}
\medskip
\vskip 0.25 true in
\centerline{Nasser Saad$^\dagger$, Richard L. Hall$^\ddagger$,  and Attila B. von Keviczky$^\ddagger$}
\bigskip
{\leftskip=0pt plus 1fil
\rightskip=0pt plus 1fil\parfillskip=0pt
\obeylines
$^\dagger$Department of Mathematics and Statistics,
University of Prince Edward Island, 
550 University Avenue, Charlottetown, 
PEI, Canada C1A 4P3.\par}

\medskip
{\leftskip=0pt plus 1fil
\rightskip=0pt plus 1fil\parfillskip=0pt
\obeylines
$^\ddagger$Department of Mathematics and Statistics, Concordia University,
1455 de Maisonneuve Boulevard West, Montr\'eal, 
Qu\'ebec, Canada H3G 1M8.\par}

\vskip 0.5 true in
%---------------------------------------------------------------------------
% Abstract
%---------------------------------------------------------------------------
\centerline{\bf Abstract}\medskip
Harrell's modified perturbation theory [Ann. Phys. {\bf 105},{ 379-406 (1977)}]
is applied and extended to obtain non-power perturbation expansions for a class of singular Hamiltonians $H=-{d^2\over dx^2}+ x^2+{A\over x^2}+{\lambda\over x^\alpha},\ (A\geq 0,\alpha>2),$
known as  generalized spiked harmonic oscillators.  The perturbation expansions developed here are valid for small values of the coupling $\lambda>0,$ and they extend the results which Harrell obtained for the spiked harmonic oscillator $A=0$. Formulas for the the excited-states are also developed.

\bigskip
\noindent{\bf PACS } 03.65.Ge
\vfil\eject
%---------------------------------------------------------------------------
% 1. Introduction
%---------------------------------------------------------------------------
\nl{\bf 1. Introduction}\medskip
This is a detailed extension of Harrell's modified perturbation theory$^\sref{\harr}$ for the class of singular potentials
$$
H=-{d^2\over dx^2}+x^2+{\lambda\over x^\alpha}\quad (\lambda>0,\alpha>2),\eqno(1.1)
$$
defined on suitable domains in the Hilbert space $L^2(0,\infty)$ with solutions satisfying Dirichlet boundary conditions. By `singular' we mean that the familiar Rayleigh-Schr\"odinger series either do not exist or do not converge. The present work, motivated by Harrell$^\sref{\harr}$ and Greenlee$^\sref{\green}$, studies a 
perturbative and variational analysis of the eigenvalues and eigenfunctions for the family of singular Hamiltonians 
$$
H=H_0+\lambda V=-{d^2\over dx^2}+x^2+{A\over x^2}+{\lambda\over x^\alpha}\quad (A\geq 0)\eqno(1.2)
$$
known as generalized spiked harmonic oscillator Hamiltonian$^{\sref{\hala-\jski}}$.  The extension lies in considering $A$ to range over all non-negative real numbers instead of non-negative integers of the type $l(l+1).$  The main results are the extensions of Harrell's perturbative expansions$^\sref{\harr}$ for the ground-state eigenvalues of the spiked harmonic oscillator Hamiltonian $A=0$. In his elegant investigation, Harrell mentioned briefly the possibility of extending his theory to the case of $A=l(l+1)$, where $l$ is the angular-momentum quantum number; however, his results mostly concern perturbation expansions for ground-state energies of the spiked harmonic oscillator Hamiltonian (1.1). There are two principal reasons for this choice: 1. The interesting Klauder phenomenon$^{\sref{\klaa-\klac}}$ occurs {\it only~} in the case $A=0$, to the effect that, for sufficiently singular potentials, the perturbation term $V$ cannot be smoothly turned off ($\lambda \rightarrow 0$) in the Hamiltonian $H=H_0+\lambda V$ to restore the free Hamiltonian $H_0$; 2.  Rayleigh-Schr\"odinger perturbation series diverge at some finite order whenever $\alpha>2$.   

Klauder's phenomenon doesn't occur$^{\sref{\eks-\deh}}$ if $A > 0.$ This is the case, for example, in $N$ dimensions with $A=(l+{1\over 2}(N-1))(l+{1\over 2}(N-3))$ and $l>0,$ or with $l = 0$ and $N\neq~1~{\rm or}~3.$ In such cases the domain of the Hamiltonian $H$ {\it is} stable under the limit $\lambda\rightarrow 0.$  However, a perturbative analysis for solutions that vanish at the origin is still interesting because of the divergence of the Rayleigh-Schr\"odinger series  at some finite order for any $\alpha>2$. We are able to conclude in the present article that the Rayleigh-Schr\"odinger series will breakdown at the order $n\geq {2\nu(\gamma-1)}$ for $\alpha>2$ where $\nu={1\over \alpha-2}$ and $\gamma=1+{1\over 2}\sqrt{1+4A}$. For example, $\alpha\geq 2\gamma$ causes the perturbation series to diverge at the first order; for $\alpha\geq\gamma+1$ the second-order perturbation will diverge, etc. These results and some others concerning the convergence of Rayleigh-Schr\"odinger series which rely heavily of the application of Kato's criterion$^{\sref{\kata-\katc}}$ will be the subject of an independent investigation. In the present paper, we concentrate on the development of non-power perturbation expansions for the Hamiltonian (1.2).

Detwiler and Klauder$^{\sref{\detw}}$, in their variational study of the spiked harmonic oscillator Hamiltonian (1.1), have shown that for $2\leq \alpha<3$ the eigenvalues are given by asymptotic series to first-order so long as $\lambda>0$. But for $\alpha> 3$, the ground-state eigenvalues are given by
$$
E_0(\lambda)=3+k\lambda^{\nu}+o(\lambda^{\nu}),
$$ 
and, for $\alpha=3,$ by
$$
E_0(\lambda)=3+k^\prime\lambda\log(\lambda)+O(\lambda),
$$ 
where $k$ and $k^\prime$ are to be determined by variational means$^{\sref{\detw}}$. 
Harrell, soon afterwards, modified the Rayleigh-Schr\"odinger series by utilizing the standard WKB-approximation technique for the lowest few orders. This proved to be quite successful, and he continued to developed a special perturbation theory, now known as `singular perturbation theory', and obtained thereby the first few terms of the perturbed $\lambda$-expansion for different values of $\alpha$. This turned out to be a non-power series expansion and in fact was of exactly the same order as that of Detwiler and Klauder$^{\sref{\detw}}$. More specifically, Harrell$^{\sref{\harr}}$ showed that the asymptotic series for the ground-state eigenvalues of the Hamiltonian (1.1) are explicitly given, for $\nu={1\over \alpha-2}$, by

\noindent for $\alpha\geq 4$: 
$$
E_0(\lambda)=
3+{4\nu^{2\nu}\Gamma(1-\nu)\over \sqrt{\pi}\Gamma(1+\nu)}
\lambda^{\nu}+ O(\lambda^{2\nu}).
$$
for $3 <\alpha< 4$:
$$
E_0(\lambda)=
3+
{4\nu^{2\nu}\Gamma(1-\nu)\over\sqrt{\pi}\Gamma(1+\nu)}
\lambda^{\nu}-
{4\nu\Gamma({3-{1\over \nu}\over 2})\over (1-\nu)\sqrt{\pi}}\lambda+ O(\lambda^{2\nu}).
$$
for $\alpha=3$:
$$
E_0(\lambda)=3-{4\over \sqrt{\pi}}\lambda\log(\lambda)-
{10c\over \sqrt{\pi}}\lambda
+ O(\lambda^2\log^2(\lambda))\quad(c= .57721~56649~\dots \hbox{ Euler's constant}).
$$
For ${5\over 2} <\alpha< 3$:
$$E_0(\lambda)=
3+
{4\nu^{2\nu}\Gamma(1-\nu)\over \sqrt{\pi}\Gamma(1+\nu)}
\lambda^{\nu}+
{2\Gamma({3-\alpha\over 2})\over \sqrt{\pi}}\lambda+ O(\lambda^{2\nu}).
$$
The paper is organized as follows. In Section~2, we briefly review the regular perturbation expansions for the Hamiltonian (1.2) and we identify the conditions under which the first- and the second-order corrections of Rayleigh-Schr\"odinger series exists. In Section~3, the main theorem used for the development of non-power perturbation expansions will be introduced and proved. In Section~4, we introduced a suitable trial wave function. In Sections~5 and 6, we extend Harrell's theory to treat the generalized spiked harmonic oscillator Hamiltonians (1.2) for the cases where the Rayleigh-Schr\"odinger series fails and thereby we show that asymptotic series for the eigenvalues of the Hamiltonian (1.2) are explicitly given by
$$
\eqalign{&\hbox{For $\alpha\geq 2(2\gamma-1)$:\quad\quad }\cr
&
\quad E_0(\lambda)=
2\gamma+
{4(\gamma-1)\nu^{4\nu(\gamma-1)}\Gamma(1-2\nu(\gamma-1))\over \Gamma(\gamma)\Gamma(1+2\nu(\gamma-1))}\lambda^{2\nu(\gamma-1)}
+ O(\lambda^{4\nu(\gamma-1)})
\cr
&\hbox{For $2\gamma<\alpha<2(2\gamma-1)$: }\cr
&
\quad E_0(\lambda)=
2\gamma+
{4(\gamma-1)\nu^{4\nu(\gamma-1)}\Gamma(1-2\nu(\gamma-1))\over \Gamma(\gamma)\Gamma(1+2\nu(\gamma-1))}\lambda^{2\nu(\gamma-1)} -
{2\nu\Gamma(\gamma-{1\over 2\nu})\over (1-2\nu(\gamma-1))~\Gamma(\gamma)}\lambda
+ O(\lambda^{4\nu(\gamma-1)})
\cr
&\hbox{For $\alpha=2\gamma$: }\cr
&
\quad E_0(\lambda)=2\gamma-{1\over (\gamma-1)\Gamma(\gamma)}\lambda\log(\lambda)+
\bigg[{-c(1+\gamma)+2\log(2(\gamma-1))\over (\gamma-1)\Gamma(\gamma)}\bigg]\lambda
+ O(\lambda^2\log^2(\lambda))\cr
&\hbox{For $\gamma+1< \alpha <2\gamma$: }\cr
&
\quad E_0(\lambda)=
2\gamma+
{2\nu^{4\nu(\gamma-1)}\Gamma(1-2\nu(\gamma-1))\over \nu\Gamma(\gamma)\Gamma(2\nu(\gamma-1))}\lambda^{2\nu(\gamma-1)} + {2\nu\Gamma(\gamma-{1\over 2\nu})\over (2\nu(\gamma-1)-1)~\Gamma(\gamma)}\lambda
+ O(\lambda^2),}
$$
where $\nu={1\over \alpha-2}$ and  $\gamma=1+{1\over 2}\sqrt{1+4A}$.
The asymptotic expansions for the case $\alpha \leq \gamma+1$ are discussed in Section~7, along with some other cases. The connection with the region $0<\alpha<5/2$, overlooked by Harrell, is also investigated.  In Section~8, the extension of the perturbation expansions developed in Sections~5, 6, and 7 to the excited states is discussed and some explicit formulas are derived.
\medskip
%----------------------------------------------------------------------------------------
%	2. Asymptotic perturbation expansions
%----------------------------------------------------------------------------------------
\nl{\bf 	2. Asymptotic perturbation expansions}\medskip
It is known that although many perturbation expansions diverge, they may actually be asymptotic expansions whose first few terms can yield good approximations. The class of singular Hamiltonian (1.2) affords interesting examples of this phenomenon. Indeed, by regarding the Gol'dman and Krivchenkov Hamiltonian$^{\sref{\hale}}$ 
$H_0=-{d^2\over dx^2}+x^2+{A\over x^2}$, which admits the exact solutions
$$\psi_n(x)=(-1)^n\sqrt{{2(\gamma)_n}\over n!\Gamma(\gamma)}x^{\gamma-{1\over 2}}e^{-{1\over 2}x^2}{}_1F_1(-n,\gamma,x^2)\eqno(2.1)
$$
with exact eigenenergies
$$E_n=4n+2\gamma,\quad n=0,1,2,\dots,\quad \gamma=1+{1\over 2}\sqrt{1+4A},\eqno(2.2)$$
as the unperturbed part, and $V(x)= x^{-\alpha}$ as the perturbation potential, the first-order correction of the Rayleigh-Schr\"odinger series  for the Hamiltonian (1.2) exists only for $\alpha<2\gamma$, while the second-order correction will required $\alpha<\gamma+1$. The first condition $\alpha<2\gamma$ follows from $E_1=(\psi_0,x^{-\alpha}\psi_0)$, while the second condition $\alpha<\gamma+1$ follows$^{\sref{\halc}}$ from  $$E_2=\sum\limits_{i=1}^\infty {|(\psi_0,x^{-\alpha}\psi_i)|^2\over E_i-E_0}.$$
Under these conditions the perturbation expansions for the ground-state eigenvalues up to second-order$^{\sref{\halc}}$ reads, for small values of $\lambda$,
$$\eqalign{
E(\lambda,\alpha)&=E_{0}+E_{1}\lambda+E_{2}\lambda^2+\dots\cr
&= 2\gamma +{\Gamma(\gamma-{\alpha\over 2})\over \Gamma(\gamma)}\lambda-\lambda^2{\alpha^2\over 16\gamma}~{}_4F_3(1,1,1+{\alpha\over 2},1+{\alpha\over 2};2,2,\gamma+1;1)+\dots.}\eqno(2.3)
$$ 
In the next sections we develop non-power perturbation expansions for the cases where the regular Rayleigh-Schr\"odinger series fails to exist; namely $\alpha\geq 2\gamma$ and $\alpha\geq \gamma+1$.

Before we proceed we should note that the functions  ${}_1F_{1}$ and ${}_4F_3$, mentioned above, are special cases of the generalized hypergeometric function$^{\sref{\luk}}$ 
$$
{}_pF_{q}(\alpha_1,\alpha_2,\dots,\alpha_p;\beta_1,\beta_2,\dots,\beta_q;z)=\sum\limits_{k=0}^\infty 
{\prod\limits_{i=1}^p(\alpha_i)_k\over  \prod\limits_{j=1}^q(\beta_j)_k}{z^k\over k!},\eqno(2.4)
$$
\nl where $p$ and $q$ are non-negative integers, and none of the $\beta_j,$ ($j=1,2,\dots,q$) is equal to zero or to a negative integer. If the series does not terminate (that is to say, none of the $\alpha_i$, $i=1,2,\dots,p$, is a negative integer), then the series, in the case $p=q+1$, converges or diverges accordingly as $|z|<1$ or $|z|>1$. For $z=1$, the series is convergent  provided 
$
{\sum\limits_{j=1}^q \beta_j-\sum\limits_{i=1}^p \alpha_i}>0.
$ 

Here $(a)_n$, the shifted factorial (or {\it Pochhammer symbol}), is defined by
$$(a)_0=1,\quad (a)_n=a(a+1)(a+2)\dots (a+n-1), \quad{\rm for}\ n = 1,2,3,\dots,\eqno(2.5)$$
\nl and may be expressed in terms of the Gamma function by $(a)_k={\Gamma(a+k)/ \Gamma(a),}$ when $a$ is not a negative integer $-m$, and, in these exceptional cases, $(-m)_k = 0$ if $k > m$ and otherwise $(-m)_k = (-1)^k m!/(m-k)!.$

\medskip
%----------------------------------------------------------------------------------------
%3.	The main theorem
%----------------------------------------------------------------------------------------
\nl{\bf 3. The Main theorem}
\medskip
It is clear that the perturbation approach mentioned in Sec. 2 cannot apply if  $\alpha\geq 2\gamma$, since it is clear in this case that the first-order perturbation correction diverges. We construct a modified perturbation series for the operators in this region by considering the perturbation theory of families of self-adjoint operators by an application of the variational method. This is done via Kato's generalization$^{\sref{\kato-\harrell2}}$ of Temple's inequality$^{\sref{\gtem-\hayes}},$ which can understood from the following discussion.
The derivation of bounds on the eigenvalues for self-adjoint operators usually starts from a consideration of the positive definite function given by
$$
(\mu,\mu)=([H-\epsilon]\phi,[H-\epsilon]\phi)=(H\phi,H\phi)-(\phi,H\phi)^2+(\epsilon-(\phi,H\phi))^2\geq 0,\eqno(3.1)
$$
where $\mu$ is a function of $\phi$ and $\epsilon$, i.e. $\mu=\mu(\phi,\epsilon)$, $H$ is the operator in question, $\epsilon$ is a positive parameter, and $\phi$ is a suitably chosen normalized trial function. If we expand the normalized function $\phi$ in terms of the complete set of eigenfunctions $\{\phi_n\}$ of $H$ with eigenvalues $E_n(\lambda)$, $\phi=\sum_{n}a_n\phi_n$, $a_n=(\phi,\phi_n)$,  $(\phi,\phi)=1=\sum_{n}|a_n|^2$, we can express the positive definite function in (3.1) as
$$
(\mu,\mu)=\sum_{n}|a_n|^2(E_n(\lambda)-\epsilon)^2\geq 0.
$$
Let us assume that we have picked the value of $\epsilon$ for the closest approach to the $i$th eigenvalue $E_i$, i.e.
$$
(\mu,\mu)=\sum_{n}|a_n|^2(E_n(\lambda)-\epsilon)^2\geq (E_i(\lambda)-\epsilon)^2 \geq 0.\eqno(3.2)
$$
By combining (3.1) and (3.2), it can be easily seen that
$$
\epsilon-\sqrt{\parallel H\phi\parallel^2 -(\phi,H\phi)^2+(\epsilon-(\phi,H\phi))^2}\leq E_i(\lambda)\leq \epsilon+\sqrt{\parallel H\phi\parallel^2-(\phi,H\phi)^2+(\epsilon-(\phi,H\phi))^2} \eqno(3.3)
$$
Now, by setting$^{\sref{\hayes}}$
$$
\epsilon+[(H\phi,H\phi)-(\phi,H\phi)^2+(\epsilon-(\phi,H\phi))^2]^{1/2}=E_{i+1}^L(\lambda),\eqno(3.4)
$$
where $E_{i+1}^L(\lambda)$ is a lower bound estimate of $E_{i+1}(\lambda)$, we can show that equation (3.4) possesses the solution
$$
\epsilon={1\over 2}\bigg[E_{i+1}^L(\lambda)+(\phi,H\phi)-{(H\phi,H\phi)-(\phi,H\phi)^2\over E_{i+1}^L(\lambda)-(\phi,H\phi)}\bigg]\eqno(3.5)
$$
provided $(\phi,H\phi)< E_{i+1}^L(\lambda)$. Substituting (3.5) into the lower bound expression in (3.3) yields the Kato-Temple expression for the lower bound:
$$
E_i(\lambda)\geq (\phi,H\phi)-{(H\phi,H\phi)-(\phi,H\phi)^2\over E_{i+1}^L(\lambda)-(\phi,H\phi)}.\eqno(3.6)
$$
Similarly, setting
$$
\epsilon-[(H\phi,H\phi)-(\phi,H\phi)^2+(\epsilon-(\phi,H\phi))^2]^{1/2}=E_{i-1}^U(\lambda),\eqno(3.7)
$$
where $E_{i-1}^U(\lambda)$ is an upper bound estimate to the next lowest eigenvalue to $E_{i}(\lambda),$ yields
$$
E_i(\lambda)\leq (\phi,H\phi)+{(H\phi,H\phi)-(\phi,H\phi)^2\over (\phi,H\phi)-E_{i-1}^U(\lambda)}\eqno(3.8)
$$
for $E_{i-1}^U(\lambda)<(\phi,H\phi)$. We let 
$\eta=(\phi,H\phi),$ and the residual norm $\varepsilon=\parallel (H-\eta)\phi \parallel$ (hence $\varepsilon^2=\parallel H\phi \parallel^2-\eta^2$), and 
$\varepsilon^2< (E_{i+1}^L(\lambda)-\eta)(\eta -E_{i-1}^U(\lambda)),$ which follows by means of the inequalities $\eta-\varepsilon^2/(E_{i+1}^L(\lambda)-\eta)> E_{i-1}^U$  or $\eta+\varepsilon^2/(\eta-E_{i-1}^U(\lambda))<E_{i+1}^L(\lambda).$ This indeed ensures that the open interval $(E_{i-1}^U(\lambda),E_{i+1}^L(\lambda))$ contains a single isolated eigenvalue and no other piece of the spectrum.  Then it follows from (3.6) and (3.8) that
$$
\eta-{\varepsilon^2\over E_{i+1}^L(\lambda)-\eta}\leq E_i(\lambda)\leq \eta+{\varepsilon^2\over \eta-E_{i-1}^U(\lambda)}.\eqno(3.9)
$$
This formula is symmetric with respect to upper and lower bound, as we might expect. It should be noted that (3.9) gives $E_i(\lambda)$ within the error bound of the order $\varepsilon^2:$ this is very small if $\varepsilon$ is small, i.e. if $\phi$ is a good approximate eigenfunction. Indeed, ~(3.9) implies 
$$|E_i(\lambda)- \eta| \leq {\varepsilon^2\over g},\eqno(3.10)
$$
where $g=\min\{\eta-E_{i-1}^U(\lambda),E_{i+1}^L(\lambda)-\eta\}$. Therefore, the error in $\eta$ depends on the residual norm squared, i.e. on $\varepsilon^2$, and on the gap $g>0$ for the eigenvalue is isolated$^\sref{\harr}$. If (3.9) is applied to the operator $H_0+\lambda V-E_i^\lambda$, where $E_i^\lambda$ is a  variational estimate for the $i^{th}$ eigenvalue of $H_0+\lambda V$, there results$^\sref{\harr}$
\medskip
\noindent{\th Theorem 1:} 
\nl
\noindent{\it If $\phi$ is normalized trial function for the self-adjoint operator $H=H_0+\lambda V$, where $H_0$ and $V$ are self-adjoint and $E_i^0$ is an isolated, nondegenerate stable eigenvalue of $H_0$, and $E_i^\lambda
$ is a continuous function such that
$(\phi,[H_0+\lambda V -E_i^\lambda]\phi)\rightarrow 0$ as $\lambda\rightarrow 0$, and
$$
\parallel [H_0+\lambda V-E_i^\lambda]\phi \parallel = o((\phi,[H_0+\lambda V-E_i^\lambda]\phi)^{1/2}),\eqno(3.11)
$$
then the eigenvalue of $H_0+\lambda V$ which converges to $E_i^0$ satisfies
$$
E_i(\lambda)=(\phi,[H_0+\lambda V]\phi)+O(\parallel [H_0+\lambda V-E_i^\lambda]\phi\parallel^2).\eqno(3.12)
$$}\medskip
\noindent {Proof:}~ To keep the notation simple, let us refer to $E_{i-1}^U(\lambda)$ and $E_{i+1}^L(\lambda)$ in (3.9) by $\alpha$ and $\beta$, respectively. Then, from the previous discussion, we have, for $\eta=(\phi,[H_0+\lambda V-E_i^\lambda]\phi),$ that
$\alpha<\eta<\beta,$ and $\varepsilon^2<(\beta-\eta)(\eta-\alpha)$. Further, by applying the Kato-Temple inequality (3.9) to the Hamiltonian $H_0+\lambda V-E_i^\lambda$, we obtain, for normalized $\phi$,
$$
\eta-{\varepsilon^2\over \beta-\eta}\leq E_i(\lambda)-E_i^\lambda\leq \eta+{\varepsilon^2\over \eta-\alpha},
$$
where $\varepsilon^2=\parallel [H_0+\lambda V-E_i^\lambda]\phi\parallel^2 - (\phi,[H_0+\lambda V-E_i^\lambda]\phi)^2$. If we divide by $\eta$, we obtain after some simplifications
$$
-{\bigg({\parallel [H_0+\lambda V-E_i^\lambda]\phi\parallel\over \eta^{1/2}}\bigg)^2-\eta\over \beta-\eta} \leq {E_i(\lambda)-E_i^\lambda\over \eta}-1\leq 
{\bigg({\parallel [H_0+\lambda V-E_i^\lambda]\phi\parallel\over \eta^{1/2}}\bigg)^2-\eta\over \eta-\alpha}.\eqno(3.13)
$$
However, since $\phi$ is assumed to be normalized,
$$
{E_i(\lambda)-E_i^\lambda\over \eta}-1=
{E_i(\lambda)-([H_0+\lambda V]\phi,\phi)\over \eta}
={E_i(\lambda)-([H_0+\lambda V]\phi,\phi)\over \parallel [H_0+\lambda V-E_i^\lambda]\phi\parallel^2}{\bigg({\parallel [H_0+\lambda V-E_i^\lambda]\phi\parallel\over \eta^{1/2}}\bigg)^2.}
$$
From (3.11), we have for $\lambda$ sufficiently small,
$$
{\parallel [H_0+\lambda V-E_i^\lambda]\phi\parallel\over \eta^{1/2}}\leq 1.$$
Thus after dividing (3.13) by $\bigg({\parallel [H_0+\lambda V-E_i^\lambda]\phi\parallel\over \eta^{1/2}}\bigg)^2$, we have
$$\eqalign{
\bigg|{E_i(\lambda)-([H_0+\lambda V]\phi,\phi)\over \parallel [H_0+\lambda V-E_i^\lambda]\phi\parallel^2}\bigg|&\leq \bigg\{ 1- {\eta\over  \bigg({\parallel [H_0+\lambda V-E_i^\lambda]\phi\parallel\over \eta^{1/2}}\bigg)^2}\bigg\}\times \max\{{1\over \beta-\eta},{1\over \eta-\alpha}\}\cr
&\leq \max\{{1\over \beta-\eta},{1\over \eta-\alpha}\}\leq C {\rm ~(constant)},
}
$$
which leads to
$$E_i(\lambda)=([H_0+\lambda V]\phi,\phi)+O(\parallel [H_0+\lambda V-E_i^\lambda]\phi\parallel^2),
$$
as required.\qed
\medskip
%----------------------------------------------------------------------------------------
%4.	Trial wave function and the differential equation
%----------------------------------------------------------------------------------------
\nl{\bf 4. Trial wave function and solution to a differential equation}
\medskip
\noindent In this section we shall introduce a suitable trial function in order to obtain eigenvalue perturbation corrections by means of Theorem 1. For singular Hamiltonians of type (1.2), the trial functions are characterized by wave functions with non-integer exponent. This indeed characterizes$^{\sref{\zno-\jam}}$ almost all trial functions which have been used previously to study this type of singular Hamiltonian (1.1) and (1.2). Furthermore, the trial functions have to satisfy the  physical initial conditions of the problem.  In the classical Rayleigh-Schr\"odinger perturbation theory, the lowest-order trial function for a given eigenvalue is chosen to be the unperturbed eigenfunction, i.e. the exact solutions of the unperturbed Hamiltonian. This is no longer a good choice for the perturbation $\lambda V$  in (1.2) with $\alpha \geq 2\gamma$, for
$$
\int\limits_{\epsilon}^\infty x^{2\gamma-\alpha-1}e^{-x^2}dx \approx \epsilon^{-\alpha+2\gamma},
$$
which approaches $\infty$ as $\epsilon$ goes to zero.
Intuitively, it seems that if the unperturbed eigenfunction was modified slightly near the singular point, so that the expectation value of singular term $V$ was no longer infinite, it would become a reasonable trial function to use to estimate the perturbed eigenvalue. This was the basic idea of the trial wave function used by Detwiler {\it et al} to study the Hamiltonian (1.1) and it was employed later by Harrell$^\sref{\harr}$. Using the notation of Harrell, we start with the (un-normalized) trial wavefunction
$$
\psi(x;\lambda)=W_\alpha (x;\lambda) \psi_i(x),\eqno(4.1)
$$
where $\psi_i(x)$ is given by (2.1) and $W_\alpha (x;\lambda)$ is to be determined.  It should be noted that far away from the singularity, we expect $\psi(x;\lambda)\sim \psi_i(x)$ for large $x$, since (1.2) behaves as radial harmonic oscillator Hamiltonian for large $x$, which, in turn, implies  $\lim\limits_{x\rightarrow \infty} W_\alpha (x;\lambda)=1.$ Further, for an arbitrary singular point $x_0$, not necessarily at the origin, $\psi(x_0;\lambda)=0$, an idea that was borrowed from hard-core problems in quantum mechanics$^{\sref{\detw}}:$ this forces $W_\alpha (x_0;\lambda)=0;$ therefore, we must also have  $\lim\limits_{x\rightarrow 0} W_\alpha (x;\lambda)=0$. Using the trial function (4.1), the differential operator (1.2) leads to
$$
[H_0+\lambda V-E_i]\psi(x;\lambda)=\bigg[-{d^2W_\alpha(x;\lambda)\over dx^2}-2{dW_\alpha(x;\lambda)\over dx}{d\over dx}+\lambda V W_\alpha(x;\lambda)\bigg]\psi_i(x), \eqno(4.2)
$$
where $E_i\equiv E_i^\lambda$ is the variational estimate of $H$. It is clear from (2.1) that ${d\psi_i(x)\over dx}\approx{\gamma - {1\over 2}\over x}\psi_i(x)$ near the origin. Therefore, we may choose $ W_\alpha(x;\lambda)$ in (4.2) such that 
$$
{d^2W_\alpha(x;\lambda)\over dx^2}+ {2(\gamma - {1\over 2})\over x} {dW_\alpha(x;\lambda)\over dx}-\lambda V W_\alpha(x;\lambda)=0\eqno(4.3)
$$
and must satisfies the initial conditions
$$\lim\limits_{x\rightarrow 0} W_\alpha (x;\lambda)=0,\hbox{ and } \lim\limits_{x\rightarrow \infty} W_\alpha (x;\lambda)=1.\eqno(4.4)$$
Eq.(4.3) allows us to write Eq.(4.2) as
$$
[H_0+\lambda V-E_i]\psi(x;\lambda)=2 {d W_\alpha(x;\lambda)\over dx}
\bigg[{(\gamma - {1\over 2})\over x}-{d\over dx}\bigg] \psi_i(x)\eqno(4.5)
$$
To solve (4.3) explicitly, we notice first that the parameter $\lambda$ can be removed from the equation by a change of variable $z=\lambda^{-\nu}x$ where $\nu$ is to be determine shortly. A straightforward calculation shows that Eq.(4.3) becomes
$$
{d^2W_\alpha(z)\over dz^2}+ {2(\gamma - {1\over 2})\over z} {dW_\alpha(z)\over dz}-{\lambda^{(2-\alpha)\nu+1}\over z^\alpha} W_\alpha(z)=0.
$$
So, with $\nu = {1\over \alpha-2}$, independent of $\gamma$, we have
$$
{d^2W_\alpha(z)\over dz^2}+ {2(\gamma - {1\over 2})\over z} {dW_\alpha(z)\over dz}-{ W_\alpha(z)\over z^\alpha}=0.\eqno(4.6)
$$
With another change of variable $Y(z) = z^{\gamma-1} W_\alpha(z)$, (4.6) leads to
$$
{d^2Y\over dz^2}+ {1\over z} {dY\over dz}-\bigg[{ (\gamma-1)^2\over z^2}+{1\over z^\alpha}\bigg] Y=0.\eqno(4.7)
$$
Finally with the further change of variable $\xi = 2\nu z^{-{1\over 2\nu}}$, we have from (4.7)
$$
{d^2Y\over d\xi^2}+ {1\over \xi} {dY\over d\xi}-\bigg[1+{ [2\nu(\gamma-1)]^2\over \xi^2}\bigg] Y=0,\eqno(4.8)
$$
which is the equation of a modified Bessel function$^{\sref{\as1}}$ of order $2\nu(\gamma-1)$. The solution of Eq.(4.8) is
$$
W_\alpha(z)=c_1~z^{1-\gamma}I_{2\nu(\gamma-1)}(2\nu z^{-{1\over 2\nu}})+c_2~z^{1-\gamma}K_{2\nu(\gamma-1)}(2\nu z^{-{1\over 2\nu}}),
$$
where $I$ and $K$ denote the modified Bessel functions of the first- and second- kind respectively$^{\sref{\as1}}$. 
The initial conditions
$
\lim\limits_{z\rightarrow 0} W_\alpha(z)=0,$ and $\lim\limits_{z\rightarrow \infty} W_\alpha(z)=1
$
yields
$c_1=0$ and $c_2={2\nu^{2\nu(\gamma-1)}\over \Gamma(2\nu(\gamma-1))}$ by means of
$$
K_{\nu}(z)\approx {1\over 2}\Gamma(\nu) {z\over 2}^\nu\eqno(4.9)
$$
as $z$ approach 0. Therefore, we have
$$\eqalign{
&W_\alpha(z)={2\nu^{2\nu(\gamma-1)}\over \Gamma(2\nu(\gamma-1))}~z^{1-\gamma}K_{2\nu(\gamma-1)}(2\nu z^{-{1\over 2\nu}}),\quad\hbox{ or more explicitly } \cr
&W_\alpha(x;\lambda)={2\nu^{2\nu(\gamma-1)}\over \Gamma(2\nu(\gamma-1))}~\lambda^{\nu(\gamma-1)}x^{1-\gamma}K_{2\nu(\gamma-1)}(2\nu \sqrt{\lambda}x^{-{1\over 2\nu}}).
}\eqno(4.10)
$$
Finally, we have for the (un-normalized) wave function (4.1) that 
$$
\psi(x;\lambda)={2\nu^{2\nu(\gamma-1)}\over \Gamma(2\nu(\gamma-1))}~\lambda^{\nu(\gamma-1)}x^{1-\gamma}K_{2\nu(\gamma-1)}(2\nu \sqrt{\lambda}x^{-{1\over 2\nu}})
\psi_i(x).\eqno(4.11)
$$
It is quite clear by means of Eq.(4.9) that $\lim\limits_{\lambda\rightarrow 0}\psi(x;\lambda)=\psi_i(x)$  as expected.  Consequently, the normalization constant $N_\lambda$ of $\psi(x;\lambda)$ must satisfy $\lim\limits_{\lambda\rightarrow 0} N_\lambda=1.$   Some properties of the function $K_\nu(z)$ are in order$^{\sref{\as1}}$. The physical importance$^{\sref{\as1}}$ of the function $K_\nu(z)$ lies in the fact that it tends exponentially to zero as $z\rightarrow \infty$. The function $K_\nu(z)$ is defined, for unrestricted values of $\nu$, by the equation
$$
K_\nu(z)={\pi\over 2\sin(\nu\pi)} [ I_{-\nu}(z)-I_\nu(z)],\eqno(4.12)													
$$
where
$$
I_\nu(z)=({z\over 2})^\nu\sum\limits_{k=0}^\infty 
 {({1\over 2}z^2)^k\over k!~\Gamma(\nu+k+1)}.\eqno(4.13)										
$$
The apparent discrepancy with (4.12) is resolved by the identity $\Gamma(\nu)\Gamma(1-\nu)={\pi\over \sin(\nu\pi)}.$ 
For integer values or zero of $\nu$ in (4.12), it should be understood that
$
K_n(z)=\lim\limits_{\nu\rightarrow n}
K_\nu(z),										
$
where in this case
$$\eqalign{
K_n(z)&={1\over 2}({z\over 2})^{-n}\sum\limits_{k=0}^{n-1}{(n-k-1)!\over k!}(-{z^2\over 4})^k+(-1)^{n+1}\log({z\over 2})I_n(z)\cr
&+(-)^n{1\over 2}({z\over 2})^{n}\sum\limits_{k=0}^{\infty}\{\psi(k+1)+\psi(n+k+1)\}
{({z^2\over 4})^k\over k!(n+k)!},								
}\eqno(4.14)
$$
while
$$
K_0(z)=-(c+\log({z\over 2}))I_0(z)+\sum\limits_{r=1}^{\infty}{({z\over 2})^{(2r)}\over (r!)^2}\bigg\{
1+{1\over 2}+{1\over 3}+\dots+{1\over r}\bigg\}.\eqno(4.15)								
$$
Here $c$ is  Euler's constant $c=.57721~56649~\dots$.
The following identity will be also in used
$$
z{dK_\nu(z)\over dz}=-\nu K_\nu(z)-zK_{\nu+1}(z).\eqno(4.16)
$$
\medskip
%----------------------------------------------------------------------------------------
%5.	Lowest-order asymptotic perturbation corrections for $\alpha\geq 2\gamma$
%----------------------------------------------------------------------------------------
\nl{\bf 5. Lowest-order asymptotic perturbation corrections for $\alpha\geq 2\gamma$}
\medskip
In this section, we apply Theorem 1 and the trial function developed in section 4 in order to obtain the eigenvalue perturbation expansions for the Hamiltonian (1.2). We consider first the case of $\alpha \geq 2\gamma$ which leads to the divergence of the first-order correction of the regular Rayleigh-Schr\"odinger series. In a purely theoretical approach, Greenlee$^{\sref{\green}}$ has shown that the asymptotic perturbation expansion should take the form
$$E_0(\lambda) = E_0 + E_1 \lambda^{2\nu(\gamma-1)}\eqno(5.1)$$
valid for $2\gamma<\alpha$ and ${2\nu(\gamma-1)}<1.$ We note, for consistency, that we have re-produced the expression of Greenlee using our own notation. Eq.(5.1) is in complete agreement with our prediction $\alpha>2\gamma$ or ${2\nu(\gamma-1)}<1$ for $\nu={1\over \alpha-2}$ obtained by means of Theorem 1, as we shall show in this section. In order to apply Theorem 1, we need first the normalization constant $N_\lambda$ of the trial wave function $\psi(x;\lambda)$, namely Eq.(4.11). This  can be found by means of the condition $\parallel \psi(x;\lambda)\parallel^2 =1$ which leads to expression
$$
N_\lambda^{-2}={4\nu^{4\nu(\gamma-1)}\lambda^{2\nu(\gamma-1)}\over [\Gamma(2\nu(\gamma-1)]^2}\int_0^\infty x^{2(1-\gamma)}\bigg[
K_{2\nu(\gamma-1)}(2\nu \sqrt{\lambda}x^{-{1\over 2\nu}})\bigg]^2\psi_i(x)^2dx.\eqno(5.2)
$$
\medskip
% ------------------------------------------------------
\noindent {\th Lemma 1:}  
%-------------------------------------------------------
{\it For the ground-state, i.e. $i=0$, we have  
$$
N_\lambda^{2}=1+{2\nu^{4\nu(\gamma-1)}\Gamma(1-2\nu(\gamma-1))\over \Gamma(\gamma)\Gamma(1+2\nu(\gamma-1))}\lambda^{2\nu(\gamma-1)}+O(\lambda^{4\nu(\gamma-1)}),\eqno(5.3)														
$$
where $\nu={1\over \alpha-2}$ and $\alpha>2\gamma$.}% end of \it
\medskip
\noindent{Proof:} We note, by using (4.9) in (5.2), that $N_\lambda^{2}\approx 1$. To find the order of the error term, however, we use the identity (4.12) which leads to $K_\nu(z)={\Gamma(\nu)\over 2}({z\over 2})^{-\nu}-{\Gamma(1-\nu)\over 2\nu}({z\over 2})^{\nu}+\dots.$ Therefore
$$\eqalign{
[K_{2\nu(\gamma-1)}(2\nu \sqrt{\lambda} x^{-{1\over 2\nu}})]^2&= {[\Gamma(2\nu(\gamma-1))]^2\over 4}(\nu\sqrt{\lambda}x^{-{1\over 2\nu}})^{-4\nu(\gamma-1)}-{\Gamma(2\nu(\gamma-1)\Gamma(1-2\nu(\gamma-1))\over 4\nu(\gamma-1)}\cr
&+{[\Gamma(1-2\nu(\gamma-1))]^2\over 4\nu^2}(\nu\sqrt{\lambda}x^{-{1\over 2\nu}})^{4\nu(\gamma-1)}+\dots.}\eqno(5.4)
$$
For the ground-state, we have from (2.1) that $\psi_0(x)= \sqrt{{2\over \Gamma(\gamma)}}x^{\gamma-{1\over 2}}e^{-x^2/2}.$ Thus on substituting (5.4) into (5.2) we have, after some calculations,
$$
N_\lambda^{2}=\bigg\{1-{2\nu^{4\nu(\gamma-1)}\Gamma(1-2\nu(\gamma-1))\over \Gamma(\gamma)\Gamma(1+2\nu(\gamma-1))}\lambda^{2\nu(\gamma-1)}+O(\lambda^{4\nu(\gamma-1)})\bigg\}^{-1}														
$$
and the proof of the lemma follows by a very similar argument to that for Taylor's expansion of $1/(1-\delta).$\qed

The reason of quoting the expansion (5.3) only up to order $\lambda^{4\nu(\gamma-1)}$ was guided by the error term in (3.12), as the following lemma indicates. 
\medskip
% ------------------------------------------------------
\noindent {\th Lemma 2:} 
%-------------------------------------------------------
{\it For the ground-state energy of the Hamiltonian (1.2), where $\alpha > 2\gamma$ (or $2\nu(\gamma-1) < 1)$, we have
$$
\eqalign{
E_0(\lambda)&=2\gamma+{16\over \Gamma(\gamma)} {\nu^{4\nu(\gamma-1)}\lambda^{2\nu(\gamma-1)+ {1\over 2}}\over [\Gamma(2\nu(\gamma-1)]^2}\cr
&\times\int\limits_0^\infty x^{1-{1\over 2\nu}}e^{-x^2}K_{2\nu(\gamma-1)}(2\nu\sqrt{\lambda}x^{-{1\over 2\nu}})
K_{1-2\nu(\gamma-1)}(2\nu\sqrt{\lambda}x^{-{1\over 2\nu}})dx + O(\lambda^{4\nu(\gamma-1)}).													
}\eqno(5.5)
$$
}\medskip
\noindent{Proof:}  Eq. (4.10) with (4.16) leads to 
$$
{dW_\alpha(x;\lambda)\over dx}={2\nu^{2\nu(\gamma-1)}\over \Gamma(2\nu(\gamma-1))}~\lambda^{\nu(\gamma-1)+{1\over 2}}x^{-{1\over 2\nu}-\gamma}K_{1-2\nu(\gamma-1)}(2\nu \sqrt{\lambda}x^{-{1\over 2\nu}}).\eqno(5.6)
$$
Furthermore, using $\psi_0(x)= \sqrt{{2\over \Gamma(\gamma)}}x^{\gamma-{1\over 2}}e^{-x^2/2}$, we find
$$\eqalign{
{d W_\alpha(x;\lambda)\over dx}
\bigg[ {(\gamma - {1\over 2})\over x}-{d\over dx}\bigg] \psi_0(x)&={2\nu^{2\nu(\gamma-1)}\lambda^{\nu(\gamma-1)+{1\over 2}}\over \Gamma(2\nu(\gamma-1))}\sqrt{{2\over \Gamma(\gamma)}}~x^{{1\over 2}-{1\over 2\nu}}e^{-x^2/2}\cr
&\times K_{1-2\nu(\gamma-1)}(2\nu \sqrt{\lambda}x^{-{1\over 2\nu}}),
}\eqno(5.7)
$$
which leads to
$$\eqalign{
2(W_\alpha(x;\lambda)\psi_0(x)&, {d W_\alpha(x;\lambda)\over dx}
\bigg[{(\gamma - {1\over 2})\over x}-{d
\over dx}\bigg] \psi_0(x))={16\over \Gamma(\gamma)}{\nu^{4\nu(\gamma-1)}\lambda^{2\nu(\gamma-1)+{1\over 2}}\over [\Gamma(2\nu(\gamma-1)]^2}\cr
&\times \int\limits_0^\infty x^{1-{1\over 2\nu}}e^{-x^2}K_{2\nu(\gamma-1)}(2\nu\sqrt{\lambda}x^{-{1\over 2\nu}})
K_{1-2\nu(\gamma-1)}(2\nu\sqrt{\lambda}x^{-{1\over 2\nu}})dx
}\eqno(5.8)
$$
\nl In order to use theorem 1, however, the trial wave function must be normalized. This is equivalent to multiplying (5.8) by the normalization constant $N_\lambda^2,$ as given by (5.3). Now, since $N_\lambda^2$ is of order $\lambda^{2\nu(\gamma-1)}$, out of the second term in (5.3) the multiplication allows us to have (5.8) as quoted, plus an error term of order $\lambda^{4\nu(\gamma-1)}$ as result of using (4.9). What remains is to show that the expression $\parallel [H_0+\lambda V-E(\lambda)]\phi_\lambda\parallel$ in (3.12) is also of order $\lambda^{2\nu(\gamma-1)}$. This follows from (5.7) as follows
$$\eqalign{
\parallel [H_0+\lambda V-E_0]\psi_0\parallel &=2
\parallel {d W_\alpha(x;\lambda)\over dx}
\bigg[ {(\gamma - {1\over 2})\over x}-{d\over dx}\bigg] \psi_0(x)\parallel\cr
&=2\bigg[{8\nu^{4\nu(\gamma-1)}\lambda^{2\nu(\gamma-1)+1}\over \Gamma(\gamma)[\Gamma(2\nu(\gamma-1)]^2} \int\limits_0^\infty x^{1-{1\over \nu}}e^{-x^2}
[K_{1-2\nu(\gamma-1)}(2\nu\sqrt{\lambda}x^{-{1\over 2\nu}})]^2dx\bigg]^{1/2}\cr
&=O(\lambda^{2\nu(\gamma-1)})
}
$$
where we have used $K_{1-2\nu(\gamma-1)}(2\nu \sqrt{\lambda} x^{-{1\over 2\nu}})\approx {\Gamma(1-2\nu(\gamma-1))\over 2}(\nu\sqrt{\lambda}x^{-{1\over 2\nu}})^{2\nu(\gamma-1)-1}$. The proof of the lemma then follows by use of Theorem 1, Eq. (3.12), and the variational estimate of $E_0^\lambda$ by means of (2.2).\qed
\medskip
Because of the error term in (5.5), it is not necessary to compute the integral in (5.5) exactly but it is sufficient to estimate the integral using the asymptotic series expansions of the modified Bessel functions $K_{2\nu(\gamma-1)}(2\nu\sqrt{\lambda}x^{-{1\over 2\nu}})$ and $
K_{1-2\nu(\gamma-1)}(2\nu\sqrt{\lambda}x^{-{1\over 2\nu}})$ by means of (4.12), up to the order cited. Since the order of the error term in (5.5) is $\lambda^{4\nu(\gamma-1)}$ while the integral is of order $\lambda^{2\nu(\gamma-1)}$, we may consider, for fixed $\alpha$, two regions $0< 2\nu(\gamma-1)\leq {1\over 2}$ and ${1\over 2} < 2\nu(\gamma-1)< 1,$ or equivalently $0< 4\nu(\gamma-1)\leq 1$ and $1 < 4\nu(\gamma-1)< 2$. For the first region, we have for the 
ground-state energy of the Hamiltonian (1.2) 
$$
E_0(\lambda)=
2\gamma+
{4(\gamma-1)\nu^{4\nu(\gamma-1)}\Gamma(1-2\nu(\gamma-1))\over \Gamma(\gamma)\Gamma(1+2\nu(\gamma-1))}\lambda^{2\nu(\gamma-1)} + O(\lambda^{4\nu(\gamma-1)})\eqno(5.9)													
$$
which follows from (5.5) using the asymptotic expansions of $K_{2\nu(\gamma-1)}(2\nu\sqrt{\lambda}x^{-{1\over 2\nu}})$ and 
$K_{1-2\nu(\gamma-1)}(2\nu\sqrt{\lambda}x^{-{1\over 2\nu}})$ by means of (4.9). In the case $\gamma=3/2$ (i.e. $A=0$ or $\alpha\geq 4$), Eq. (5.9) yields
$$
E_0(\lambda)=
3+
{4\nu^{2\nu}\Gamma(1-\nu)\over \sqrt{\pi}\Gamma(1+\nu)}\lambda^{\nu} + O(\lambda^{2\nu}),\eqno(5.10)												
$$
\nl where $\nu = {1\over \alpha-2}$, as shown earlier by Harrell for the spiked harmonic oscillator Hamiltonian (1.1). Important conclusion follows from (5.9). For $\alpha= 2(2\gamma-1)$ or $2\nu(\gamma-1)={1\over 2}$, we have
$$
E_0(\lambda)=
2\gamma+
{2\over \Gamma(\gamma)}\sqrt{\lambda} + O(\lambda).\eqno(5.11)													
$$
\nl This provides a single ground-state approximation formula for a wide class of Hamiltonians $H=-{d^2\over dx^2}+x^2+{A\over x^2}+{\lambda\over x^\alpha},$ where
$\alpha$ and $\gamma=1+{1\over 2}\sqrt{1+4A}$ are related by $\alpha= 2(2\gamma-1)$. For example, for $A=0, i.e.~\gamma=3/2,$ which yields $\alpha=4$, we have  
$$
E_0(\lambda)=
3+
{4\over \sqrt{\pi}}\sqrt{\lambda} + O(\lambda),													
$$
\nl as noted by Harrell.  If $\alpha = 6,$ which implies $\gamma=2$~ or $A = 0.75$, we have
$$
E_0(\lambda)=
4+
2\sqrt{\lambda} + O(\lambda).												
$$

For the second region $1 < 4\nu(\gamma-1)< 2$, or $2\gamma<\alpha<2(2\gamma-1)$, by using (4.12), we can easily show that
$$\eqalign{
K_{2\nu(\gamma-1))}(2\nu \sqrt{\lambda} x^{-{1\over 2\nu}})&K_{1-2\nu(\gamma-1))}(2\nu \sqrt{\lambda} x^{-{1\over 2\nu}})=
{\Gamma(2\nu(\gamma-1))\Gamma(1-2\nu(\gamma-1))\over 2\nu \sqrt{\lambda}}
x^{{1\over 2\nu}}\cr
&-{[\Gamma(2\nu(\gamma-1))]^2\over 4(1-2\nu(\gamma-1))}\nu^{1-4\nu(\gamma-1)}\lambda^{{1\over 2}-2\nu(\gamma-1)} x^{-{1\over 2\nu}+2(\gamma-1)}+\dots.}
$$
\nl Consequently, (5.5) yields, for ${1\over 2} < 2\nu(\gamma-1)< 1$,
$$
E_0(\lambda)=
2\gamma+
{4(\gamma-1)\nu^{4\nu(\gamma-1)}\Gamma(1-2\nu(\gamma-1))\over \Gamma(\gamma)\Gamma(1+2\nu(\gamma-1))}\lambda^{2\nu(\gamma-1)} -
{2\nu\Gamma(\gamma-{1\over 2\nu})\over (1-2\nu(\gamma-1))~\Gamma(\gamma)}\lambda
+ O(\lambda^{4\nu(\gamma-1)}).\eqno(5.12)													
$$
\nl Again the result of Harrell for the Hamiltonian (1.1) follows for the case of $\gamma={3\over 2}$, i.e. $A=0$, where, in this case,  $3 < \alpha< 4$ or ${1\over 2}<\nu<1$, and
$$
E_0(\lambda)=
3+
{4\nu^{2\nu}\Gamma(1-\nu)\over \sqrt{\pi}\Gamma(1+\nu)}
\lambda^{\nu}-
{4\nu\Gamma({3-{1\over \nu}\over 2})
\over (1-\nu)\sqrt{\pi}}\lambda+ O(\lambda^{2\nu}).\eqno(5.13)												
$$
For the rest of this section, we consider the case of $2\nu(\gamma-1)=1$. For this specific value Eqs. (4.10) and (5.6) read, for $z=\lambda^{-\nu}x$,
$$
W_\alpha(z)={z^{1-\gamma}\over \gamma-1}K_{1}({z^{1-\gamma}\over \gamma-1})\eqno(5.14)
$$
\nl and
$$
{dW_\alpha(z)\over dz}={z^{1-2\gamma}\over \gamma-1}K_{0}({z^{1-\gamma}\over \gamma-1}
),\eqno(5.15)
$$
respectively. Using the asymptotic expansions
$$
K_0(z)=[-c+\log(2)-\log(z)]+O(z^2),\quad\quad				
K_1(z)={1\over z}+O(z)\eqno(5.16)						
$$
which follow by means of (4.15) and (4.14) respectively, we obtain
\medskip
% ----------------------------------------------
\noindent {\th Lemma 3:} 
% ----------------------------------------------
{\it For the ground state energy of the Hamiltonian (1.2), where $\alpha= 2\gamma$ (or  $2\nu(\gamma-1)=1)$, we have
$$
E_0(\lambda)=2\gamma-{1\over (\gamma-1)\Gamma(\gamma)}\lambda\log(\lambda)+
\bigg[{-c(1+\gamma)+2\log(2(\gamma-1))\over (\gamma-1)\Gamma(\gamma)}\bigg]\lambda
+ O(\lambda^2\log^2(\lambda)),	\eqno(5.17)										
$$
where $c = .57721~56649~\dots$ is Euler's constant.}
\medskip
% ------------------------------------------------------
\noindent{Proof:} We should note first, in this case,
% ------------------------------------------------------
$$
\parallel [H_0+\lambda V-E(\lambda)]\psi_0\parallel^2 =O(\lambda^2\log^2(\gamma))
$$
which follows from
$$
\parallel [H_0+\lambda V-E(\lambda)]\psi_0\parallel^2 ={2\lambda^2\over (\gamma-1)^2\Gamma(\gamma)}\int_0^\infty x^{3-2\gamma}e^{-x^2}K_0^2({\sqrt{\lambda}\over \gamma-1}x^{1-\gamma})dx
$$
by use of (5.7). Since we are only interested in finding the order in terms of the parameter $\lambda$, the problem reduces to a search among the smallest value of $\lambda^2\log^2(\lambda)$, $-\lambda^2\log(\lambda)$, and $\lambda^2,$ for small values of the parameter $\lambda$. Therefore
for sufficiently small $\lambda$ we have $
\parallel [H_0+\lambda V-E(\lambda)]\psi_0\parallel^2 =O(\lambda^2\log^2(\lambda)) 
$  as noted . What remains is to compute 
$$\eqalign{
2(W_\alpha(x;\lambda)\psi_0(x), &{d W_\alpha(x;\lambda)\over dx}
\bigg[ {(\gamma - {1\over 2})\over x}-{d\over dx}\bigg] \psi_0(x))
={4\lambda^{3/2}\over (\gamma - 1)^2\Gamma(\gamma)}\cr
&\times\int_0^\infty x^{2 - \gamma}e^{-x^2}K_0({\sqrt{\lambda}\over \gamma-1}x^{1-\gamma})K_1({\sqrt{\lambda}\over \gamma-1}x^{1-\gamma})dx
}
$$
by using the asymptotic expansions (5.16) up to the order $\lambda^2\log^2(\lambda)$. The lemma then follows after some straightforward calculations. It is important to note that the normalization constant $N_\lambda,$ as given by (5.2), yields in this case
$$
N_\lambda^{-2}=1+\bigg\{2\nu^2(-1+2c+2\log(\nu))+{1\over 2}\nu c\bigg\} {\lambda\over \Gamma(\gamma)}+{2\nu^2\over \Gamma(\gamma)}\lambda\log\lambda+\dots
$$
and will contribute to the error term in a similar manner to that mentioned in lemma 2.\qed
\medskip
The results of Harrell, the case $\alpha = 3$, follows immediately from (5.17) for the case of $A=0$ (or $\gamma=3/2$), i.e. $\nu = 1$, namely 
$$
E_0(\lambda)=3-{4\over \sqrt{\pi}}\lambda\log(\lambda)-
{10c\over \sqrt{\pi}}\lambda
+ O(\lambda^2\log^2(\lambda)).\eqno(5.18)	
$$
It is clear that these expressions are valid for $\lambda$ much smaller than unity. 
\medskip
%----------------------------------------------------------------------------------------
%6.	Lowest-order asymptotic perturbation corrections for $\alpha\geq \gamma+1$
%----------------------------------------------------------------------------------------
\nl{\bf 6. Lowest-order asymptotic perturbation corrections for $2\gamma>\alpha\geq \gamma+1$}
\medskip
In this section, we discuss the case of $2\gamma>\alpha\geq \gamma+1$ or equivalently the case of $1< 2\nu(\gamma-1)\leq 2$. It is clear by now that, for $1< 2\nu(\gamma-1)\leq 2$, the first-order Rayleigh-Schr\"odinger corrections exist but the second-order corrections diverge. Thus, the improved perturbation procedure gives an explicit term between the first and the second order. Let us first consider the case of $1< 2\nu(\gamma-1)< 2$, here we rely on the asymptotic expansion of the modified Bessel function $K$ as given by (4.12). We note from (4.5) and (5.6) that
$$
\parallel [H_0+\lambda V-E(\lambda)]\psi_0\parallel^2 ={32\nu^{4\nu(\gamma-1)}\lambda^{2\nu(\gamma-1)+1}\over \Gamma(\gamma)[\Gamma(2\nu(\gamma-1))]^2}~\int_0^\infty x^{1-{1\over \nu}}e^{-x^2}
K_{2\nu(\gamma-1)-1}^2(2\nu \sqrt{\lambda}x^{-{1\over 2\nu}})dx
$$
as a consequence of the known identity $K_\nu(z)=K_{-\nu}(z)$. Using (4.9) we have
$$
\parallel [H_0+\lambda V-E(\lambda)]\psi_0\parallel^2 =O(\lambda^2)\quad\quad (1< 2\nu(\gamma-1)< 2),\eqno(6.1)
$$
which leads to the following lemma.
\medskip
\noindent {\th Lemma 4:} 
{\it For the ground-state energy of the Hamiltonian (1.2) where $\alpha> \gamma+1$ (i.e. $1<2\nu(\gamma-1)<2)$, we have
$$\eqalign{
E_0(\lambda)&=2\gamma+{16\over \Gamma(\gamma)}{\nu^{4\nu(\gamma-1)}\lambda^{2\nu(\gamma-1)+{1\over 2}}\over [\Gamma(2\nu(\gamma-1)]^2}\cr
&\times\int\limits_0^\infty x^{1-{1\over 2\nu}}e^{-x^2}K_{2\nu(\gamma-1)}(2\nu\sqrt{\lambda}x^{-{1\over 2\nu}})
K_{2\nu(\gamma-1)-1}(2\nu\sqrt{\lambda}x^{-{1\over 2\nu}})dx + O(\lambda^2).}
\eqno(6.2)													
$$
}\medskip
\noindent The proof of this lemma is similar to that of lemma 3, therefore we omit it.
The computation of the integral in (6.2) up to the order of $\lambda^2$ yields the perturbation expansion 
$$
E_0(\lambda)=
2\gamma+
{2\nu^{4\nu(\gamma-1)}\Gamma(1-2\nu(\gamma-1))\over \nu\Gamma(\gamma)\Gamma(2\nu(\gamma-1))}\lambda^{2\nu(\gamma-1)} + {2\nu\Gamma(\gamma-{1\over 2\nu})\over (2\nu(\gamma-1)-1)~\Gamma(\gamma)}\lambda
+ O(\lambda^2),\eqno(6.3)													
$$
as the result of  
$$\eqalign{
K_{2\nu(\gamma-1)}&(2\nu \sqrt{\lambda} x^{-{1\over 2\nu}})= {\Gamma(2\nu(\gamma-1))\Gamma(1-2\nu(\gamma-1))\over 2}\bigg\{(\nu\sqrt{\lambda}x^{-{1\over 2\nu}})^{-2\nu(\gamma-1)}\bigg[{1\over \Gamma(1-2\nu(\gamma-1))}\cr
&+{(\nu \sqrt{\lambda} x^{-{1\over 2\nu}})^2\over \Gamma(2-2\nu(\gamma-1)}\bigg]
- (\nu\sqrt{\lambda}x^{{1\over 2\nu}})^{2\nu(\gamma-1)}\bigg[{1\over \Gamma(1+2\nu(\gamma-1))}+{(\nu \sqrt{\lambda} x^{-{1\over 2\nu}})^2\over \Gamma(2+2\nu(\gamma-1)}\bigg]\bigg\}+\dots.
}
$$ 
\nl The result of Harrell$^{\sref{\harr}}$ follows immediately from (6.3) in the special case $A=0$, namely
$$
E_0(\lambda)=
3+
{4\nu^{2\nu}\Gamma(1-\nu)\over \sqrt{\pi}\Gamma(1+\nu)}
\lambda^{\nu}+
{2\Gamma({3-\alpha\over 2})
\over \sqrt{\pi}}\lambda+ O(\lambda^{2\nu}),\eqno(6.4)												
$$
where $\nu = {1\over \alpha-2}$ and $5/2<\alpha<3$.

For the case $2\nu(\gamma-1)=2$, the norm $\parallel [H_0+\lambda V-E(\lambda)]\psi_0\parallel$ can be computed easily by means of Eq.(4.5), which yields 
$$\eqalign{
\parallel [H_0+\lambda V-E_0]\psi_0\parallel^2 &=2
\parallel {d W_\alpha(x;\lambda)\over dx}
\bigg[ {(\gamma - {1\over 2})\over x}-{d\over dx}\bigg] \psi_0(x)\parallel^2\cr
&={16\lambda^{3}\over (\gamma-1)^2\Gamma(\gamma)} \int\limits_0^\infty x^{2-\gamma}e^{-x^2}
[K_{1}({2\sqrt{\lambda}\over \gamma-1}x^{-{1\over 2\nu}})]^2dx=O(\lambda^{2}),
}
$$
as a consequence of  
$$
K_1(z)={1\over z}+({1\over 4}(-1 + 2 c) + {1\over 2}(-\log(2) + \log(z)))z+O(z^2).
$$
Furthermore, since the trial wave function takes the form
$$
\psi(x;\lambda)=2\nu^2\lambda x^{-{1\over \nu}}K_2(2\nu \sqrt{\lambda}x^{-{1\over 2\nu}})\psi_0(x),
$$
we may compute the normalization constant Eq.(5.2) as
$$
N_\lambda^2=1+{\nu\Gamma({1\over 2\nu})\over \Gamma(\gamma)}\lambda+{\nu^4\over \Gamma(\gamma)}\lambda^2\log(\lambda)+O(\lambda^2).
$$
in a similar fashion to the proof of lemma 1. 
Harrell, in his investigation, claims that the ground-state eigenvalue perturbation expansion for the spiked harmonic oscillator Hamiltonian (1.1), $\alpha=5/2,$ is given by
$$
E_0(\lambda)=3+{2\Gamma({1\over 4})\over \sqrt{\pi}}\lambda+{16\over \sqrt{\pi}}\lambda^2\log(\lambda)+O(\lambda^2).\eqno(6.5)			
$$
The following result does not confirm his claim, but shows that it is slightly different even by means of Harrell's own methodology. As our calculation will show, the ground-state perturbation expansion in the case of $\alpha=5/2$ is actually given by
$$
E_0(\lambda)=3+{2\Gamma({1\over 4})\over \sqrt{\pi}}\lambda+{32\over \sqrt{\pi}}\lambda^2\log(\lambda) + O(\lambda^2),\eqno(6.6)
$$
\nl with a multiple of 2 in the log term in contrast with (6.5).
In order to verify (6.6), we adopt two different approaches, first we use Harrell's method then we apply our generalization. We have for $\alpha=5/2$, the asymptotic perturbation expansion according to Harrell reads
$$
E_0(\lambda)=3+2(W_{5/2}\psi_0,{dW_{5/2}\over dx}({1\over x}-{d\over dx})\psi_0)+O(\lambda^2),
$$
where $\psi_0(x)={2\over \pi^{1/4}}xe^{-x^2/2}$. For $W_{5/2}$, Harrell used the asymptotic approximation of the modified Bessel function 
$$
K_2(z)={2\over z^2}-{1\over 2}+({1\over 16}({3\over 2} - 2c) +{1\over 8}(\log(2) - \log(z)))z^2+O(z^3)
$$
to show that
$$
W_{5/2}(z)=1-{4\over \sqrt{z}}+{4\log(z)\over z}-4(4(c+\log(2))-3)z+\dots\quad\hbox{for large z}.
$$
\nl If we differentiate $W_{5/2}(z)$ with respect to $x$, keeping in mind $z=\lambda^{-2}x$, we have, after some calculations,
$$
2(W_{5/2}\psi_0,{dW_{5/2}\over dx}({1\over x}-{d\over dx})\psi_0)={16\lambda\over \sqrt{\pi}}\int\limits_0^\infty x^{3/2}e^{-x^2}dx+{64\lambda^2\log{\lambda}\over \sqrt{\pi}}\int\limits_0^\infty x e^{-x^2}dx+\dots,
$$
which yields (6.6), since $\int\limits_0^\infty x e^{-x^2}dx={1\over 2}$ and $\int\limits_0^\infty x^{3/2}e^{-x^2}dx={1\over 8}\Gamma({1\over 4})$. Numerically, however, Eq.(6.5) is more appealing than (6.6) for a wider range of the parameter $\lambda$ smaller than unity since (6.6) reduces the applicable range of $\lambda$ by almost one-half. 
\medskip
%--------------------------------------------------------------------
\noindent {\th Lemma 5:} 
%--------------------------------------------------------------------
{\it For the ground-state energy of the Hamiltonian (1.2), where $ 2\nu(\gamma-1)= 2$, we have
$$
E_0(\lambda)=2\gamma+{\Gamma({1\over 2\nu})\over \Gamma(\gamma)}\lambda+{2\nu^3\over \Gamma(\gamma)}\lambda^2\log(\lambda)+O(\lambda^2).\eqno(6.7)			
$$}
\medskip
\nl{Proof:} For $2\nu(\gamma-1)=2$, we have, using Eq.(6.2), that
$$
E_0(\lambda)=2\gamma+{16\nu^{4}\lambda^{{5\over 2}}\over\Gamma(\gamma)}
\int\limits_0^\infty x^{1-{1\over 2\nu}}e^{-x^2}K_{2}(2\nu\sqrt{\lambda}x^{-{1\over 2\nu}})
K_{1}(2\nu\sqrt{\lambda}x^{-{1\over 2\nu}})dx + O(\lambda^2).													
$$
By means of the asymptotic expansions
$$
K_1(z)={1\over z}+({1\over 4}(-1+2c)+{1\over 2}\log({1\over 2}z))z+O(z^2)
$$
\nl and 
$$
K_2(z)={2\over z^2}-{1\over 2}+({1\over 16}({3\over 2}-2c)-{1\over 8}\log({1\over 2}z)){z^2}
+
O(z^4),
$$
\nl where $z=2\nu\sqrt{\lambda}x^{-{1\over 2\nu}}$, we have, after some calculations up to order $\lambda^2$, that
$$
E_0(\lambda)= 2\gamma+{4\nu\lambda\over \Gamma(\gamma)}
\int\limits_0^\infty x^{1+{1\over \nu}}
e^{-x^2}dx+
{4\nu^3\over \Gamma(\gamma)} \lambda^2\log(\lambda)
\int\limits_0^\infty xe^{x^2} dx+O(\lambda^2).												
$$
\nl This leads to (6.7), since $\int\limits_0^\infty x^{1+{1\over \nu}}
e^{-x^2}dx={1\over 2}\Gamma(1+{1\over 2\nu})$.\qed
\medskip
%--------------------------------------------------------------
\nl{\bf 7. Further Cases}
% -------------------------------------------------------------
\medskip
The expansions developed above can be extended to the case of $2\nu(\gamma-1)>2$ and the region of $\alpha< {5\over 2}$ which was not studied by Harrell$^{\sref{\harr}}$. For example, in the case of  $2< 2\nu(\gamma-1)<3$, the second order of the perturbation correction exists but the third order diverges. By using the modified perturbation theory we find
$$\eqalign{
E_0(\lambda)=
2\gamma&+
{4(\gamma-1)\nu^{4\nu(\gamma-1)}\Gamma(1-2\nu(\gamma-1))\over \Gamma(\gamma)\Gamma(1+2\nu(\gamma-1))}\lambda^{2\nu(\gamma-1)} +
{2\nu\Gamma(\gamma-{1\over 2\nu})\over (2\nu(\gamma-1)-1)~\Gamma(\gamma)}\lambda\cr
&-{2(3-4\nu(\gamma-1))\nu^3\Gamma(\gamma-{1\over \nu})\over (2-2\nu(\gamma-1))(1-2\nu(\gamma-1))^2\Gamma(\gamma)}\lambda^2+O(\lambda^3).									
}\eqno(7.1)
$$
\nl Again in the case of $\gamma=3/2$, Harrell's formula should read, for $7/3<\alpha<5/2$,
$$
E_0(\lambda)=
3+
{4\nu^{2\nu}\Gamma(1-\nu)\over \sqrt{\pi}\Gamma(1+\nu)}\lambda^{\nu} +
{4\nu\Gamma({3\over 2}-{1\over 2\nu})\over (\nu-1)~\sqrt{\pi}}\lambda-{4(3-2\nu)\nu^3\Gamma({3\over 2}-{1\over \nu})\over (2-\nu)(1-\nu)^2\sqrt{\pi}}\lambda^2+O(\lambda^3),\eqno(7.2)								
$$
where $\nu={1\over \alpha-2}$. Furthermore, for  $3< 2\nu(\gamma-1)<4$, we have
$$\eqalign{
E_0(\lambda)=
2\gamma&+
{4(\gamma-1)\nu^{4\nu(\gamma-1)}\Gamma(1-2\nu(\gamma-1))\over \Gamma(\gamma)\Gamma(1+2\nu(\gamma-1))}\lambda^{2\nu(\gamma-1)} +
{2\nu\Gamma(\gamma-{1\over 2\nu})\over (2\nu(\gamma-1)-1)~\Gamma(\gamma)}\lambda\cr
&-{2(3-4\nu(\gamma-1))\nu^3\Gamma(\gamma-{1\over \nu})\over (2-2\nu(\gamma-1))(1-2\nu(\gamma-1))^2\Gamma(\gamma)}\lambda^2-{2\nu^5\Gamma(\gamma-{3\over 2\nu})\over \Gamma(\gamma)(1-2\nu(\gamma-1))^2(2-2\nu(\gamma-1))}\lambda^3\cr
&+O(\lambda^4).}\eqno(7.3)									
$$

\nll Similar expressions can be obtained for $n<2\nu(\gamma-1)<n+1,~n=4,5,\dots,$ as well for $2\nu(\gamma-1)= n$. It is important, however, to note that there are an infinite number of cases as $n$ increase to infinity for $\alpha>2$. For example, if we restrict the value of $\gamma$ to $3/2$, then in this case $2<\alpha<5/2,$ and obtain an infinite number of perturbation expansions consisting of analytic parts of degree $\lambda^n$ in addition to one correction term in each expansion as $2+{1\over 1+n}<\alpha<2+{1\over n}$. Similarly, the cases $\alpha=2+{1\over 1+n},~n = 3,4,\dots,$  need special treatment, as we have mentioned above. It is interesting to note that the \hi{\lambda}{term} in (7.1) or in (7.3) is identical with the corresponding \hi{\lambda}{term}  in the Rayleigh-Schr\"odinger series. This can be easily verified by a comparison of the coefficient of $\lambda$ in (7.1) or in (7.3) with the coefficient of the \hi{\lambda}{term} in (2.3). However this is clearly not the case for the \hi{\lambda^2}{term}. 
\medskip
%--------------------------------------------------------------
\nl{\bf 8. Excited states}
% -------------------------------------------------------------
\medskip
The perturbation expansions developed so far were restricted to the ground-state energy, however, it is a matter of calculation to extend these results to the excited-state energies. First, it should be noted that the exact solution of the unperturbed part of the Hamiltonian (1.2) has very little to do with the order of the error terms in the perturbation expansions. Therefore it is expected that the order of the error terms remains the same for the excited states, i.e. $i=1,2,\dots$. Following the discussion in Sec. 3, the asymptotic expansions for the eigenvalues for excited states are given by means of Theorem 1 as
$$\eqalign{
E_i(\lambda)&=2(2i+\gamma)+2(W_\alpha(x;\lambda)\psi_i(x), {d W_\alpha(x;\lambda)\over dx}
\bigg[{(\gamma - {1\over 2})\over x}-{d
\over dx}\bigg] \psi_i(x))\cr
&+O(\parallel [H_0+\lambda V-E_i^\lambda]W_\alpha(x;\lambda)\psi_i\parallel^2),
}\eqno(8.1)$$
\nl where $\psi_i(x),\quad i=0,1,2,\dots$  are given by (2.1) and the energy of the unperturbed Hamiltonian $E_i^\lambda$ is given by (2.2). In order to compute (8.1) explicitly, we notice first that
$$
\eqalign{
2(W_\alpha(x;\lambda)\psi_i(x),& {d W_\alpha(x;\lambda)\over dx}
\bigg[{(\gamma - {1\over 2})\over x}-{d
\over dx}\bigg] \psi_i(x))
=
{16 (\gamma)_i\over i!\Gamma(\gamma)}
{\nu^{4\nu(\gamma-1)}\lambda^{2\nu(\gamma-1)+{1\over 2}} \over [\Gamma(2\nu(\gamma-1)]^2}
\cr
&\times\int\limits_0^\infty x^{1-{1\over 2\nu}}e^{-x^2}\bigg[{2(\gamma+i)\over \gamma}{}_1F_1(-i;\gamma;x^2){}_1F_1(-i;\gamma+1;x^2)-[{}_1F_1(-i;\gamma;x^2)]^2\bigg]\cr
&\times K_{2\nu(\gamma-1)}(2\nu\sqrt{\lambda}x^{-{1\over 2\nu}})K_{1-2\nu(\gamma-1)}(2\nu\sqrt{\lambda}x^{-{1\over 2\nu}}),
dx
}\eqno(8.2)$$
\nl where ${}_1F_1$ is the confluent hypergeometric function mentioned earlier Eq.(2.4). In producing (8.2) we have used the following identity$^\sref{{\as1}}$
$${d\over dz} {}_1F_1(a;b;z)={}_1F_1(a;b;z)-{(b-a)\over b}{}_1F_1(a;b+1;z).$$
\nl For the case of $0<2\nu(\gamma-1)\leq {1\over 2}$ or $\alpha \geq 2(2\gamma-1)$, we find, by means of (4.9), that
$$
\eqalign{
2(W_\alpha(x;\lambda)\psi_i(x),& {d W_\alpha(x;\lambda)\over dx}
\bigg[{(\gamma - {1\over 2})\over x}-{d
\over dx}\bigg] \psi_i(x))
=
{4 (\gamma)_i\over i!\Gamma(\gamma)}
{\Gamma(1-2\nu(\gamma-1))\over \nu\Gamma(2\nu(\gamma-1))}
{\nu^{4\nu(\gamma-1)}\lambda^{2\nu(\gamma-1)}}
\cr
&\times\int\limits_0^\infty x e^{-x^2}\bigg[{2(\gamma+i)\over \gamma}{}_1F_1(-i;\gamma;x^2){}_1F_1(-i;\gamma+1;x^2)-[{}_1F_1(-i;\gamma;x^2)]^2\bigg]
dx.
}\eqno(8.3)$$
\nl This result required the investigation of some integrals of the type
$$
\int\limits_0^\infty x e^{-x^2}{}_1F_1(-i;\gamma;x^2){}_1F_1(-i;\gamma+1;x^2) dx
\hbox{ and }
\int\limits_0^\infty x e^{-x^2}[{}_1F_1(-i;\gamma;x^2)]^2dx.
$$
\medskip
\noindent {\th Lemma 6:} {\it
For $d > 0$ and $s > 0$
$$\int\limits_0^\infty t^{d-1}e^{-st} {}_1F_1(a;b;k t){}_1F_1(a^\prime;b^\prime;k^\prime t) dt=
s^{-d}\Gamma(d)\sum\limits_{m=0}^\infty {(a)_m(d)_m\over (b)_m}{({k\over s})^m\over m!} {}_2F_1(a^\prime,d+m;b^\prime;{k^\prime\over s}).
$$
}\medskip
\noindent{Proof:} From the series representation, Eq.(2.4), of the confluent hypergeometric series ${}_1F_1$, namely
$${}_1F_1(a;b;k t)=\sum\limits_{m=0}^\infty {(a)_m\over (b)_m}{(kt)^m\over m!}\quad\hbox{and }\quad {}_1F_1(a^\prime;b^\prime;k^\prime t)=\sum\limits_{n=0}^\infty {(a^\prime)_n\over (b^\prime)_n}{(k^\prime t)^n\over n!},$$
we have
$$\eqalign{
\int\limits_0^\infty t^{d-1}e^{-st} {}_1F_1(a;b;k t){}_1F_1(a^\prime;b^\prime;k^\prime t) dt&=
\sum\limits_{m=0}^\infty \sum\limits_{n=0}^\infty{(a)_m(a^\prime)_n\over (b)_m(b^\prime)_n}
{k^m{k^\prime}^n\over m! n!}\int\limits_0^\infty e^{-st}t^{d+n+m-1}dt\cr
&= s^{-d}\sum\limits_{m=0}^\infty \sum\limits_{n=0}^\infty{(a)_m(a^\prime)_n\over (b)_m(b^\prime)_n}
{k^m{k^\prime}^n\over m! n!}({1\over s})^n ({1\over s})^m (d+m)_n \Gamma(d+m)\cr
&=s^{-d}\sum\limits_{m=0}^\infty \bigg[\sum\limits_{n=0}^\infty
{(a^\prime)_n(d+m)_n\over (b^\prime)_n}{({k^\prime\over s})^n\over n!}
\bigg] {(a)_m({k\over s})^m\over (b)_m} \Gamma(d+m)\cr
&=s^{-d} \Gamma(d)\sum\limits_{m=0}^\infty {}_2F_1(a^\prime,d+m;b^\prime;{k^\prime\over s}) {(a)_m(d)_m\over (b)_m}{({k\over s})^m\over m!}.\quad\quad
}$$
\nl with $d > 0$ and $ s > 0$, and where we have used the series representation of ${}_2F_1$ as given by Eq(2.4) and the Pochhammer's identity $\Gamma(d+m+n)=(d+m)_n\Gamma(d+m)= (d+m)_n(d)_m\Gamma(d)$.\qed
\medskip
\nl As a consequence of this lemma, we have 
$$
\int\limits_0^\infty x e^{-x^2}{}_1F_1(-i;\gamma;x^2){}_1F_1(-i;\gamma+1;x^2)dx={(\gamma)_i\over 2(\gamma+1)_i}{}_3F_2(-i,1-\gamma,1;\gamma,1-\gamma-i;1)\eqno(8.4)
$$
\nl and 
$$\int\limits_0^\infty x e^{-x^2}[{}_1F_1(-i;\gamma;x^2)]^2dx={(\gamma-1)_i\over 2(\gamma)_i}{}_3F_2(-i,-\gamma,1;\gamma,2-\gamma-i;1),\eqno(8.5)
$$
\nl where the Chu-Vandermonda identity
$${}_2F_1(-n,a;c;1)={(c-a)_n\over (c)_n}$$
has been used. Therefore, from (8.3), we have
$$
\eqalign{
2(W_\alpha&(x;\lambda)\psi_i(x), {d W_\alpha(x;\lambda)\over dx}
\bigg[{(\gamma - {1\over 2})\over x}-{d
\over dx}\bigg] \psi_i(x))
=
{4 (\gamma)_i\over i!\Gamma(\gamma)}
{\Gamma(1-2\nu(\gamma-1))\over \nu\Gamma(2\nu(\gamma-1))}
{\nu^{4\nu(\gamma-1)}\lambda^{2\nu(\gamma-1)}}
\cr
&\times\bigg[{}_3F_2(-i,1-\gamma,1;\gamma,1-\gamma-i;1)
-{(\gamma-1)_i\over 2(\gamma)_i}{}_3F_2(-i,-\gamma,1;\gamma,2-\gamma-i;1)
\bigg]
}\eqno(8.6)$$
\nl For $0<2\nu(\gamma-1)\leq {1\over 2}$ or $\alpha \geq 2(2\gamma-1)$, we have
$$\eqalign{
E_i(\lambda)&=2(2i+\gamma)+{4 (\gamma)_i\over i!\Gamma(\gamma)}
{\Gamma(1-2\nu(\gamma-1))\over \nu\Gamma(2\nu(\gamma-1))}
{\nu^{4\nu(\gamma-1)}\lambda^{2\nu(\gamma-1)}}
\cr
&\times\bigg[{}_3F_2(-i,1-\gamma,1;\gamma,1-\gamma-i;1)
-{(\gamma-1)_i\over 2(\gamma)_i}{}_3F_2(-i,-\gamma,1;\gamma,2-\gamma-i;1)
\bigg]
\cr
&+O(\lambda^{4\nu(\gamma-1)}),\quad i=0,1,2,\dots.
}\eqno(8.7)$$
\nl Similar expressions can be obtained for the other cases by means of lemma 6. An immediate extension of Harrell's expansions to excited states $i=1,2,\dots$ for the Hamiltonian (1.1) can be obtained by setting $\gamma={3\over 2}$ in (8.7).\bigskip
% -----------------------------
\nl{\bf 9. Conclusions}\medskip
% -----------------------------
In this article, we have applied and extended Harrell's modified perturbation theory to treat a wider class of singular Hamiltonians given by (1.2). Our extensions allow us to recover Harrell's formulas for the spiked harmonic oscillator Hamiltonian (1.1) as special cases. Further, we were able to extend Harrell's results to the exited-state energies, again as special cases of our general treatment. We have also now corrected the perturbation expansion (6.5) for the case $\alpha={5\over 2}:$ this formula has been used without correction since the very early work of Harrell. Some interesting questions which remain to be answered are as follows. Is the modified wave function (4.11) sufficient to extend the perturbation expansions presented here to higher orders, or will further modifications need to be introduced?  Why can the second-order corrections in (7.1), (7.2), and (7.3) not be recovered from the corresponding terms in the regular Rayleigh-Schr\"odinger series?  Does this fact indicate that the trial wave function indeed requires additional modification?  We hope that the present work will encourage further research into this interesting class of singular Hamiltonians.
\bigskip
\noindent {\bf Acknowledgment}
\medskip Partial financial support of this work under Grant Nos. GP3438 and GP249507 from the 
Natural Sciences and Engineering Research Council of Canada is gratefully 
acknowledged by two of us (respectively [RLH] and [NS]). 

\vfil\eject

%-------------------------------------
\references{1}
%-------------------------------------

\end